# Structural and insulator-to-metal phase transition at 50 GPa in GdMnO$_3$

J. Agostinho Moreira,[1*] A. Almeida,[1] J. Oliveira,[1] V. H. Rodrigues,[2] M.M.R. Costa,[2] P. B. Tavares,[3] P. Bouvier,[4] M. Guennou,[4] J. Kreisel [4*]

[1]IFIMUP and IN-Institute of Nanoscience and Nanotechnology, Departamento de Física e Astronomia, Faculdade de Ciências, Universidade do Porto. Rua do Campo Alegre, 687, 4169-007 Porto, Portugal.
[2]CEMDRX, Departamento de Física, Faculdade de Ciências e Tecnologia, Universidade de Coimbra, 3004-516 Coimbra, Portugal.
[3]Centro de Química. Universidade de Trás-os-Montes e Alto Douro. Apartado 1013, 5001-801. Vila Real. Portugal.
[4]Laboratoire des Matériaux et du Génie Physique, CNRS, Grenoble Institute of Technology, MINATEC, 3 parvis Louis Néel, 38016 Grenoble, France.

We present a study of the effect of very high pressure on the orthorhombic perovskite GdMnO$_3$ by Raman spectroscopy and synchrotron x-ray diffraction up to 53.2 GPa. The experimental results yield a structural and insulator-to-metal phase transition close to 50 GPa, from an orthorhombic to a metrically cubic structure. The phase transition is of first order with a pressure hysteresis of about 6 GPa. The observed behavior under very high pressure might well be a general feature in rare-earth manganites.

Rare-earth manganites have attracted a continuous attention for their complex correlation between lattice, electric and magnetic degrees of freedom. More recently, magnetoelectric and multiferroic properties of manganites have attracted a particular interest.[1] Most manganites crystallize at ambient conditions in a *Pnma* structure, which presents distortions away from the ideal cubic perovskite structure through cooperative Jahn-Teller distortion and tilt of the MnO$_6$ octahedra. Such manganites have been extensively studied as a function of temperature, magnetic field, strain (in thin films), or chemical composition. High pressure is another parameter allowing tuning different degrees of freedom in manganites, but remains little explored to date. A notable exception is the study of the crystal structure, Jahn-Teller distortion, orbital order, and pressure-induced insulator-metallic phase transition in LaMnO$_3$, although the driving mechanism still remains controversial.[2-8] According to the pioneer work by Loa *et al*[2], the Jahn-Teller effect and the concomitant orbital ordering are suppressed above 18 GPa. The system retains insulator behavior up to 32 GPa, undergoing a bandwidth driven insulator-metal phase transition. Other authors have reported the persistence of the Jahn-Teller distortion over the entire stability range of the insulating phase of LaMnO$_3$, suggesting a non classical Mott insulator.[3,6]
No other orthorhombic manganite has attracted such an attention, although we note the pressure investigation of structural properties of the magnetoelectric manganites TbMnO$_3$ and DyMnO$_3$ or BiMnO$_3$, or more complex solid solutions.[9-11] To the best of our knowledge, manganites have not yet been investigated in the very high-pressure regime around 50 GPa, despite promising studies on similar orthoferrites or BiFeO$_3$, which have revealed intriguing insulator-to-metal or structural transitions in a similar pressure range.[12-15]

The aim of this work is to explore the effect of very high-pressure on rare-earth manganites. We have chosen GdMnO$_3$ which currently attracts a considerable attention as a frustrated magnetic system for which a ferroelectric order can be induced by application of a modest magnetic field.[16-18] Its phase diagram has been extensively studied as a function of external parameters: Temperature, doping, and high magnetic fields.[16,18,19] Pressure has only been explored up to 1 GPa through a pressure-dependent study of the dielectric constant at low temperatures.[20] Here, we report a pressure-dependent investigation of the lattice dynamics and crystal structure of GdMnO$_3$ at room temperature up to 53.2 GPa, through Raman spectroscopy and x-ray powder diffraction.

High quality ceramic GdMnO$_3$ samples were prepared with the sol-gel urea combustion method (see Ref. 21) and its chemical, morphological and structural characteristics were checked by x-ray, SEM, EDS and XPS. The powder sample was loaded in a DAC with diamond tips of diameter 300 μm and with helium as a pressure-transmitting medium. The Raman spectra were recorded on a LabRam spectrometer using a He-Ne laser at 633 nm. The laser power was kept below 5 mW on the DAC to avoid sample heating. High-pressure synchrotron x-ray diffraction experiments were performed at the European Synchrotron Radiation Facility (ESRF) on the ID27 high pressure beam line. X-ray diffraction patterns were collected on an image plate detector with a focused monochromatic beam with λ = 0.3839 Å. The powder diffraction data were analyzed by full Rietveld refinements using the FullProf software.

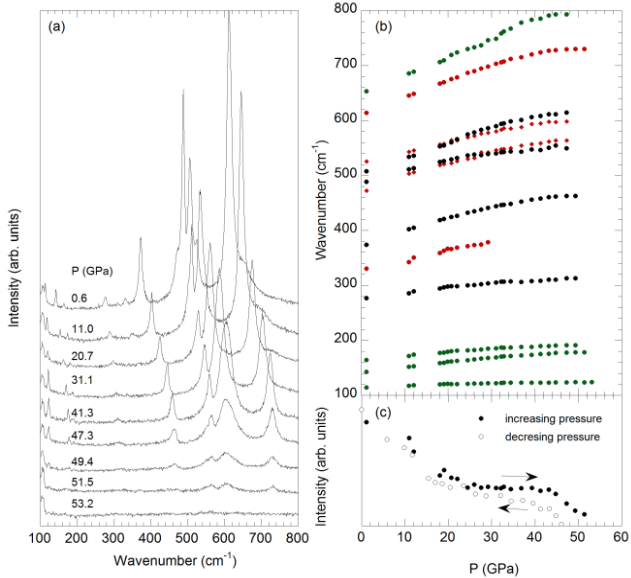

FIG. 1. (a) Evolution of the Raman spectra and (b) Raman shifts for all bands with increasing pressure. (c) Intensity of the Raman mode at initially 610 cm$^{-1}$ upon increasing and decreasing pressure.

Rare-earth manganites, which crystallize in the P*nma* structure, present 24 Raman active modes.[22] In our unpolarized Raman set-up, all Raman-active modes can be observed. The Raman spectra of GdMnO$_3$ have been studied by several authors both at low temperatures and under high magnetic field.[22-24] Lavèrdiere et al[23] proposed that the intense Raman bands in rare-earth manganites are of $A_g$ and $B_{2g}$ symmetry, which was confirmed by a study of Y-doped EuMnO$_3$.[25] Iliev et al[22] presented the mode assignment of various rare-earth manganites together with the atomic motions involved in each mode

Fig. 1(a) presents selected Raman spectra of GdMnO$_3$ for an increasing pressure up to 53.2 GPa, recorded at room temperature. The spectrum at ambient conditions is consistent with literature data (see Refs. 22,23) and is characterized by 12 bands, with prominent features at 610 cm$^{-1}$ ($B_{2g}$ Jahn-Teller symmetric stretching mode), 502 cm$^{-1}$ ($A_g$ bending mode), 487 cm$^{-1}$ ($A_g$ Jahn-Teller asymmetric stretching mode), 373 cm$^{-1}$ ($A_g$ mode, tilt of the MnO$_6$ octahedra around [101]), and 275 cm$^{-1}$ ($A_g$ mode, tilt of the MnO$_6$ octahedra around [010]). The two shoulders at 473 cm$^{-1}$ and at 522 cm$^{-1}$ correspond to bending of MnO$_6$ octahedra and scissor-like oxygen rotations modes of symmetry $B_{2g}$, respectively.

With increasing pressure, all Raman bands shift to higher wavenumbers, as expected for pressure-induced bond shortening. It can be seen from Fig. 1(a) that the overall spectral signature is maintained up to around 51 GPa. Within this pressure range, the changes in the spectral signature can be entirely explained by increasing line width, band overlap and a decreasing intensity, suggesting that GdMnO$_3$ undergoes no phase transition up to 51 GPa. This observation is supported by the pressure-evolution of the Raman shifts for the different bands, which show different slopes and curvatures but no anomaly (Fig. 1(b)). In the range from 0.1 to 25 GPa, the wavenumber of each band can be approximated by a linear function of pressure, with slopes presented in Table I. We note that two pairs of modes at 473/487 cm$^{-1}$ and 502/522 cm$^{-1}$ cross upon increasing pressure, which is allowed by their different symmetries.

| $\omega_o$ (cm$^{-1}$) | Symmetry | Slope (cm$^{-1}$/GPa) |
|---|---|---|
| 114 | | 0.03(3) |
| 141 | $B_{2g}$ or $B_{3g}$ | 0.96(3) |
| 163 | $B_{2g}$ or $B_{3g}$ | 0.78(4) |
| 275 | $A_g$ | 1.06(5) |
| 326 | $B_{2g}$ | 1.9(1) |
| 373 | $A_g$ | 2.47(8) |
| 473 | $B_{2g}$ | 2.4(1) |
| 487 | $A_g$ | 2.02(4) |
| 502 | $A_g$ | 2.82(8) |
| 522 | $B_{2g}$ | 1.88(6) |
| 610 | $B_{2g}$ | 3.10(3) |
| 648 | | 3.31(9) |

Table I: Frequencies of the experimentally Raman lines of GdMnO$_3$ at room conditions, their symmetry and the slope of the pressure-dependence of the frequency of the observed Raman bands, for the 0 – 20 GPa pressure range.

The most notable result of Fig. 1(a) is the complete extinction of the Raman signature at 53.2 GPa. As an example, Fig. 1(c) shows the pressure dependence of the intensity of the band associated with the in-phase O2 stretching mode at initially 610 cm$^{-1}$, which decreases gradually and eventually vanishes above 51.5 GPa. This suppression is a clear sign for a phase transition at very high-pressure. The transition is reversible, as illustrated through the recovery with decreasing pressure of the intensity of the in-phase O2 stretching mode as shown in Fig. 1(c). The Raman spectrum is only recovered after decreasing pressure just below 47 GPa, indicating an important pressure hysteresis of 6 GPa, which strongly suggests a first order phase transition.

While the above signature provides conclusive evidence for a phase transition, the evolution of the Raman spectra alone does not allow elucidating its nature. Two main scenarios can be envisaged: (*i*) GdMnO$_3$ shows under high-pressure a phase transition from the *Pnma* to the ideal perovskite *Pm-3m* structure. In this scenario, the loss of the Raman spectra is explained by the fact that Raman scattering is forbidden by symmetry in the ideal perovskite structure. The choice of this structure is unique since Raman scattering is not forbidden for any other perovskite structure. (*ii*) GdMnO$_3$ undergoes under high-pressure a transition from an insulator to a metal (possibly accompanied by a change in the magnetic order), likewise to what has been observed in a similar pressure range for orthoferrites or BiFeO$_3$.[12-14] Here, the loss of the Raman spectrum is explained by the enhancement of absorption coefficient in the metallic phase. This fact in turn inhibits the observation of a signal, which is already weak for the black samples in the insulator phase. (*iii*) A combination of (*i*) and (*ii*), i.e. a transition towards a metallic cubic *Pm-3m* phase, is of course also possible.

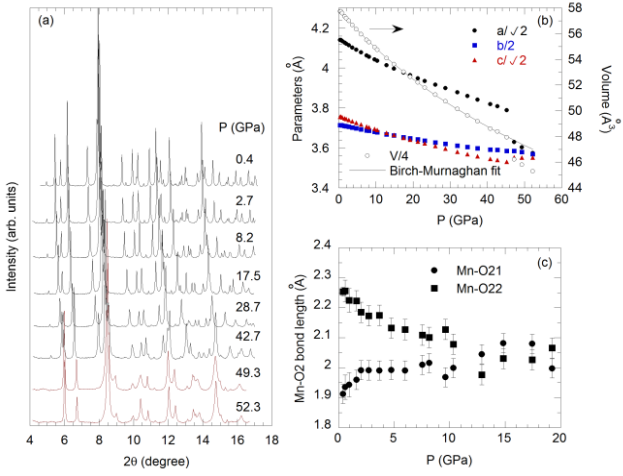

FIG. 2. Evolution of the (a) integrated diffraction pattern, (b) pseudo-cubic parameters and volume (c) Mn-O2 bond lengths under pressure. The solid line in Figure 2(b) was calculated from the best fit of the Birch-Murnaghan equation to the volume data.

In order to elucidate which of the models can best explain the observed transition, we have undertaken a pressure-dependent synchrotron x-ray diffraction experiment up to 52.3 GPa. Figure 2(a) shows representative powder x-ray pattern collected at room temperature for different pressures. The inspection of the diffraction data shows that $GdMnO_3$ maintains the characteristic diffraction pattern of a $Pnma$ structure up to 45 GPa. Above this pressure the diffraction pattern changes significantly towards a pattern with fewer Bragg reflections, suggesting an increase in symmetry. Most Bragg reflections of the new high-pressure phase are significantly larger than in the low pressure phase.

| Pressure (GPa) | Atom | Wyckoff position | x | y | z |
|---|---|---|---|---|---|
| 0.4 | Gd | 4c | 0.5812(4) | 0.250 | 0.5203(6) |
| | Mn | 4c | 0.00 | 0.00 | 0.500 |
| | O1 | 4b | -0.033(4) | 0.250 | 0.394(4) |
| | O2 | 8d | 0.164(4) | 0.548(2) | 0.196(3) |
| 52.3 | Gd1 | 4a | -0.021(1) | -0.021(1) | -0.021(1) |
| | Gd2 | 4a | 0.516(1) | 0.516(1) | 0.516(1) |
| | Mn1 | 4a | 0.277(1) | 0.277(1) | 0.277(1) |
| | Mn2 | 4a | 0.739(3) | 0.739(3) | 0.739(3) |
| | O1 | 12b | 0.340(6) | 0.310(8) | 0.014(7) |
| | O2 | 12b | 0.297(6) | 0.235(8) | 0.469(5) |

Table II: Wyckoff positions, atomic coordinates and agreement factors obtained from the Rielveld refinement of the x-ray spectra recorded at 0.4 GPa and 52.3 GPa.

In order to discuss the structural evolution in the low-pressure phase, we have carried out full Rietveld refinements of all diffraction patterns up to a pressure of 45 GPa by assuming a $Pnma$ space group, which allows a reliable fit of the whole diffraction pattern. Figure 3(a) illustrates the quality of a representative refinement at a pressure of 0.4 GPa. The analysis of the diffraction pattern obtained above 45 GPa is less straightforward, due to a phase coexistence with broad overlapping reflections and an unusual pattern. By investigating the high-pressure pattern through a profile matching, a cubic metric appears as a good candidate. However, the Raman-inactive $Pm\text{-}3m$ structure of the ideal cubic perovskite or centered cubic structures cannot explain the pattern, as we observe a number of superstructure reflections representative of a primitive structure with doubled parameters.

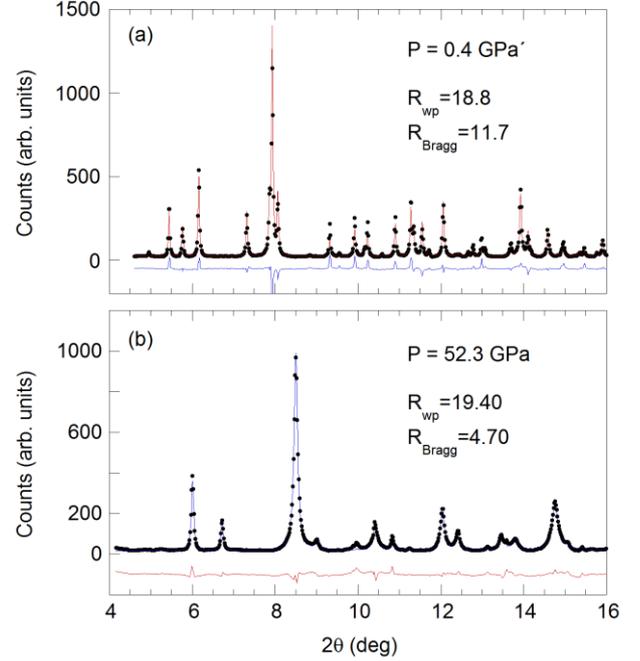

FIG. 3. Experimental, calculated and difference XRD spectra (a) at 0.4 GPa in the P$nma$ phase and (b) at 52.3 GPa in the P$2_1$3 phase.

The best profile matching is obtained by assuming $a_{cubic} = b_{Pnma}$ with a large primitive cell with $V_{cubic} = 2V_{Pnma}$ and $Z = 8$. Based on this metric, full refinements were tested for all primitive cubic space groups to identify the most appropriate model. The best fit to the experimental data was achieved for space group $P2_13$ (see profile and reliability factors in Fig. 3(b)). Table II presents the Wyckoff positions, the atomic coordinates and agreement factors resulting from the Rietveld refinement at 0.4 GPa and at 52.3 GPa. Let us parenthesize that: *(i)* this cubic cell and this space group are rather unusual for simple perovskites; *(ii)* a factor group analysis of $GdMnO_3$ in the $P2_13$ space group predicts 117 optical modes at the Γ-point of the Brillouin zone, with $\Gamma_{optical} = 10A + 10E + 29F$. As the A and E modes are Raman-active, the suppression of the Raman spectrum cannot be explained by a structural phase transition but more likely by an insulator-metallic phase transition.

The pressure dependence of pseudo-cubic lattice parameters $a/\sqrt{2}$, $b/2$, $c/\sqrt{2}$ and volume V/4 of the unit cell are shown in Fig. 2(b). Within the orthorhombic phase, the compression is anisotropic with the soft direction along the $a$-axis. We note that the order of the $b$ and $c$ lattice parameters is inversed with increasing pressure, consistent with the continuous evolution from a Jahn-Teller distorted structure at low pressure to a $GdFeO_3$-type tilted perovskite at high-pressure. The pressure dependence of the unit cell volume is adequately described by a third-order Birch-Murnaghan equation[26] up to 45 GPa, yielding $B_o = (175 \pm 2)$ GPa and $B_o' = 3.01 \pm 0.06$ for the bulk modulus and its pressure derivative at zero pressure, respectively (see solid line in Fig. 2(b)). The value of $B_o'$ is marginally outside of the range of typical values for nearly isotropic compression ($B_o' \approx 4 - 6$), reflecting anisotropy in compressibility. The 3,6% drop of V/4 is again a signature of a first-order transition. We note that the phase transition pressure observed in Raman and XRD experiments differs by some 2 GPa, which is explained by the first order nature of the phase transition and the different coherence length of Raman and XRD.

We now discuss the evolution with pressure of Mn-O bond lengths, deduced from the atomic positions obtained from Rietveld refinement of the x-ray data. The analysis is restricted to data obtained up to 20 GPa. At higher pressures the broadening of Bragg reflections limits the reliability of the structural data obtained from full Rietveld analysis, in particular the positions of the light oxygen atoms. Fig. 2(c) shows that the two initially different Mn-O2 distances approach with increasing pressure and tend to merge into a single distance at around 12 GPa. This observation evidences that the cooperative Jahn-Teller distortion is reduced with increasing pressure. Though the quality of high-pressure data does not enable to discuss when and whether the Jahn-Teller distortion is entirely suppressed, it is worth noting the decrease around 12 GPa occurs below the structural phase transition around 50 GP.

In summary, a pressure-dependent study of $GdMnO_3$ by using Raman spectroscopy and synchrotron source-ray diffraction, up to 53.2 GPa yields a phase transition around 50 GPa from an orthorhombic to a cubic structure. Its first order nature is evidenced by phase coexistence, volume change at the transition and pressure hysteresis of 6 GPa. Raman and XRD allow identifying the phase transition as being both a structural and an insulator-metallic phase transition. The refinement of pressure-data in the *Pnma* structure suggests a decrease of the Jahn-Teller distortion. This distortion is not detectable above ~12 GPa, which is well below the phase transition pressure. It is worthwhile to note that this transition occurs in the very high-pressure regime, similarly to what has been observed in rare-earth orthoferrites or $BiFeO_3$.[12-15] This behavior might well be a general feature in rare-earth manganites under very high-pressure.

This work was supported by Fundação para a Ciência e Tecnologia, through the Projects PTDC/CTM/67575/2006 and PEst-C/FIS/UI0036/2011. We would like to thank V. Komshenko for helpful discussions.